\documentclass[superscriptaddress,twocolumn,prb,showkeys,showpacs]{revtex4}

\usepackage{graphics}
\usepackage{graphicx}
\usepackage{amsmath}	
\usepackage{amsfonts}

\begin{document}

\title{Spin Dependent Joule Heating due to Rashba Coupling and Zitterbewegung }
\date{\today}

\author{Z. Wilamowski}
\affiliation{Institute of Physics, Polish Academy of Sciences, PL 02-668 Warsaw, Poland}
\affiliation{Faculty of Mathematics and Computer Sciences, UWM Olsztyn, 10-561 Olsztyn, Poland}
\author{W. Ungier}
\affiliation{Institute of Physics, Polish Academy of Sciences, PL 02-668 Warsaw, Poland}
\author{M. Havlicek}
\affiliation{Institut f. Halbleiter- und Festk\"orperphysik, Johannes Kepler Universit\"at, A-4040 Linz, Austria}
\author{W. Jantsch}
\affiliation{Institut f. Halbleiter- und Festk\"orperphysik, Johannes Kepler Universit\"at, A-4040 Linz, Austria}

\begin{abstract}
Investigating microwave absorption in asymmetric Si quantum wells in an external magnetic field, we discover a spin dependent component of Joule heating at spin resonance. We explain this effect in terms of Rashba spin-orbit coupling which results in a current induced spin precession and Zitterbewegung. Evidence is based on the observation of a specific dependence of the electron spin resonance line shape and its amplitude on the experimental geometry which in some range suggests a ``negative'' differential power absorption.
\end{abstract}



\maketitle

\def\rf{{\it r.f. }}
\def\ie{{\it i.e.}}
\def\eg{{\it e.g. }}
\def\chio{\chi^{\!\,'}}
\def\chid{\chi^{\!\,'\!\,'}}
\def\pri#1{#1^{\!\,'}}
\def\dpri#1{#1^{\!\,'\!\,'}}
\def\i{\mathrm{i}}
\def\deg{^\circ}
\def\bv#1{{\mathbf #1}}
\def\d{\mathrm{d}}
\def\myRe{\mathfrak{Re}}
\def\myIm{\mathfrak{Im}}

\section{INTRODUCTION}

In solids with lower than mirror symmetry, the Rashba-type of spin-orbit coupling for charge carriers in semiconductors and metals\cite{1,2}  gives rise, e.g., to Dyakonov-Perel spin relaxation, the spin Hall effect, electric dipole spin resonance or a group of spin photo-voltaic effects.\cite{3,4} In this paper we present evidence for another effect originating from Rashba coupling. We show that it opens a new channel of spin dependent energy transfer between an \rf electric field and the kinetic energy reservoir.
 
Rashba coupling has recently received attention also since, making use of spin-orbit (S-O) interaction, efficient spin manipulation may be achieved by exciting spin precession by an \rf electric field. This mechanism can be much more efficient\cite{5}  (one of the main preconditions for spintronics\cite{6,7,8}) than the classical excitation by an oscillating magnetic field which is used in standard electron spin resonance, ESR. Also, addressing nanostructures via electric currents is much easier than the generation of an \rf magnetic field in a small volume. 

Spin excitation via the S-O field appears well understood.\cite{9} For low frequency, the excitation efficiency is determined by the electric current, \ie, it scales with the electron mobility ({\bf{C}}urrent {\bf{I}}nduced {\bf{ESR}}).\cite{10} For high frequency, when the displacement current exceeds the drift component, momentum dissipation can be neglected and spin-flip transitions can be treated within the formalism of electric dipole transitions. Within this limit, this effect has been described many years ago by Rashba and it is called {\bf{E}}lectric {\bf{D}}ipole {\bf{ESR}} ({\bf{ED ESR}}).\cite{11}
  
The magnitude of the S-O field has been well documented in Si quantum wells.\cite{12} There is, however, a rich and hardly understood dependence of the CI ESR line shape and its anisotropy on experimental conditions\cite{12} in Si and also in other materials.\cite{13,14} In particular, all models based on the effective S-O field predict that the ESR line shape can be described by the imaginary-, \ie, the so-called absorption component of the dynamic magnetic susceptibility, $\chid(\omega)$---both for the S-O component and for any superposition of S-O---and magnetic \rf fields.\cite{9,10} This conclusion is in contrast with observation: the power absorption spectra are composed of absorption- and dispersion-like components\cite{10} of $\chi(\omega)$. Moreover, the amplitude of the dispersive component (proportional to $\myRe(\chi) = \chio(\omega)$), can be much bigger than the absorptive one. This effect is puzzling also because $\chio(\omega)$   changes its sign in some frequency range which seems to suggest the possibility of a  ``negative power absorption''.

In this paper, we investigate the ESR line shape for {\bf both} electric and magnetic microwave excitation of 2D electrons in SiGe/Si/SiGe quantum wells and we propose a theoretical explanation based on the fact that the S-O field results in a dependence of the electron velocity on spin orientation. Consequently, excitation of spin precession leads to oscillating components of the electron velocity. The interference of this motion with the \rf electric field modifies the electric absorption in the sample contributing thus to the power absorption spectrum. This effect is proportional to the precession amplitude and thus it appears at ESR. This specific type of electrical absorption can be treated thus as arising from a spin dependent electric conductivity or as a resonant modification of the Joule heat. In agreement with the experimental conditions, we model the effect within the linear response limit assuming small precession excitation, where saturation effects in the spin polarization can be neglected. In particular, we omit higher order effects, such as a dependence of the electric conductivity on spin polarization, \ie, the so-called polarization signal.\cite{12}

\section{EXPERIMENTAL}

We investigate conduction electron ESR in a one-sided, modulation doped Si quantum well defined by Si$_{0.75}$Ge$_{0.25}$ barriers as described in Refs. 5, 12. The same type of samples, grown by molecular beam epitaxy on relaxed Si$_{0.75}$Ge$_{0.25}$ layers, is used here. We use a Bruker ElexSys E580 ESR spectrometer in the X-band at temperatures of 2--40\,K. Two types of microwave cavities were employed: the standard one (TE$_{201}$) where sample is placed in the maximum of the magnetic microwave field, $H_1$, and a special cavity (TE$_{301}$) where the sample is located in a maximum of the electric field, $E_1$. In both cases 
$\bv{E_1}$ is parallel- and $\bv{H_1}$  perpendicular to the external static magnetic field, $\bv{H}$. The sample can be rotated about the direction of $\bv{H_1}$, which allows to change the direction of  $\bv{H}$ with respect to the sample growth direction, $\hat{\bv{n}}$, described an angle $\theta$. The microwave bridge works in such a way that the observed signal monitors the power absorption spectrum only. Because of the modulation of the external field, the spectrum is obtained as the first derivative of the power absorption. 

\section{EXPERIMENTAL RESULTS}

The ESR due to the 2D conduction electrons in Si produces a very narrow line (typically $<< 1$\,G) superimposed on a broad spectrum of electrical absorption.\cite{12} The latter is caused by the excitation of coupled plasmon-cyclotron oscillations and harmonics.\cite{15,16} Here the spectra are rather complex, showing up to 4 resonances\cite{17}  in the accessible range of 0--1.5\,T. The width of these resonances allows to estimate the momentum relaxation time. Results spread in the range of  $\tau_p=(3\cdot 10^{-12}$--$4\cdot10^{-11})$\,s.

\begin{figure}
\includegraphics[width=9.5cm]{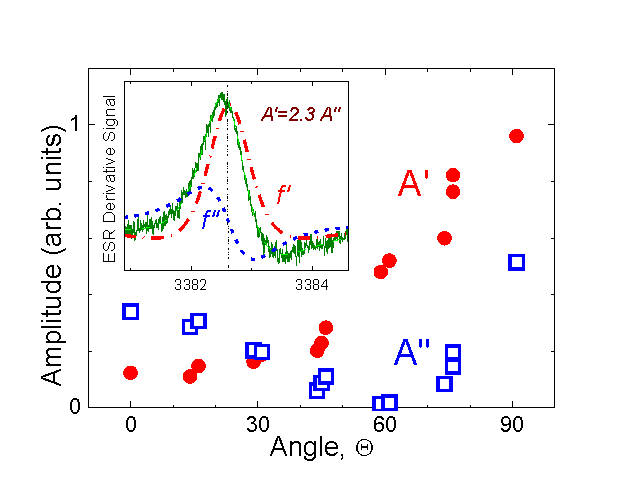}
\caption{\label{fig1} Angular dependence of the amplitudes of the real, $\pri{A}$ (red dots), and the imaginary part, $\dpri{A}$  (blue squares), of the ESR signal at $\omega=2\pi\cdot 9.4\,\mbox{GHz}$ of a high mobility 2D electron gas ($\omega\tau_p=3$) when the sample is placed at the {\bf node of $\bv{E_1}$}. Inset: observed derivative of microwave absorption ESR line (solid green line). The line shape is fitted (dashed green line) by a combination of real, $\pri{f}(B)$  (dash-dotted red), and imaginary, $\dpri{f}(B)$  (dotted blue), Lorentzians (Eq.~\ref{e1}). In this example, $\pri{A}$ is by a factor 2.3 bigger than $\dpri{A}$. }
\end{figure}

A typical example for the ESR line shape is shown in the inset of Fig.~\ref{fig1}. It is very far from the classical absorption line (dashed line) but it can be well fitted by a sum $f(H)=\pri{A}\pri{f}(H)+\dpri{A} \dpri{f}(H)$ of imaginary and real parts derived from the Lorentzian shape function:
\begin{equation} f_\pm(\omega)=\pm\frac{\i}{\pi}\frac{T_2}{1-\i(\omega\mp\omega_{\mathrm{L}})T_2}=\pri{f_\pm}(\omega)+\i\dpri{f_\pm}(\omega).
\label{e1}
\end{equation}

Here $T_2$ is the transverse spin relaxation time and $\omega_{\mathrm{L}}=\gamma H_{\mathrm{ESR}}$  the Larmor frequency at ESR, and $\gamma$  is the gyromagnetic factor. For the ESR line shown in the inset of Fig.~\ref{fig1}, the amplitude of the dispersion component, $\pri{A}$, is by a factor $2.3$ bigger than  $\dpri{A}$.  Both amplitudes, and their ratio change from sample to sample. They depend also on the direction of the applied field, $\theta$, as shown in Figs.~\ref{fig1} and \ref{fig2}.

The data in Fig.~\ref{fig1} were obtained with the sample positioned in the node of $\bv{E_1}$. The classical ESR absorption (induced by $H_1$ ) is well evidenced only for small $\theta$  by a dominant absorption amplitude, $\dpri{A}$  (see Fig.~\ref{fig1}).  Here, close to its node, $\bv{E_1}$  is small and roughly perpendicular to the sample layer. Therefore almost no current is induced. Consequently all contributions caused by the electric field are small, and the absorption signal originates from magnetic dipole (MD) transitions only. 

The MD ESR signal is expected to be independent of the sample orientation. But when $\theta$  is increased, an in-plane electric current is induced. Accordingly, $\dpri{A}$  should increase, but it decreases. Obviously another signal appears, compensating part of the amplitude of MD absorption. Simultaneously, the amplitude of the dispersive component, $\pri{A}$, substantially increases. Both components scale roughly like $(E_1\sin{\theta})^2$, indicating that both are caused by the S-O field.\cite{18}

Placing the sample in the node of $H_1$, we obtain the angular dependencies of the signal amplitudes shown in Fig.~\ref{fig2}.  In that case, for $\theta\rightarrow 0$  both amplitudes practically vanish and increase with $\theta$. Obviously, in that case the amplitude of the MD resonance is negligible and all observed ESR signals are excited by $E_1\sin{\theta}$   via S-O coupling. 	The inset shows the derivative of CR absorption spectra, \ie, the dynamic electric conductivity, measured for $\theta=0\deg, 30\deg, 60\deg$ and $90\deg$. The width of CR for that sample indicates a comparatively low mobility. With increasing $\theta$, the CR moves towards higher field, roughly with $1/\cos{\theta}$, demonstrating the dependence of CR on the perpendicular component of the static magnetic field, $H\cos{\theta}$.

\section{MODELING AND DISCUSSION}
 
The occurrence of a dispersive component in the absorption spectra (which sometimes is even negative) cannot be explained by existing models,\cite{9,10} neither for MD ESR nor ED ESR, nor for CI ESR. But the experimental data clearly indicate that the effect is induced by the \rf electric field, demonstrating that it is related to S-O coupling.
 
We consider here a system lacking mirror symmetry. In that case, a zero-field spin splitting occurs, described by the Rashba\cite{9,10}  term,  $\mathcal{H}=a_{\mathrm{R}}(s_xk_y-s_yk_x)$, or equivalently, by the S-O field, $H_{\mathrm{R}}$, acting on individual electrons with a momentum $\hbar\mathrm k$: 
\begin{equation}
g\mu_{\mathrm{B}}\mu_0\bv{H}_{\mathrm{R}}\cdot\bv{s}=a_{\mathrm{R}}(\bv{k}\times\bv{n})\cdot\bv{s}, 
\label{e2}
\end{equation}
where $\bv{s}$ stands for its spin, $a_{\mathrm{R}}$ is the Bychkov-Rashba coefficient, $g$ the g-factor, $\mu_{\mathrm{B}}$  the Bohr magneton and $\mu_0$   is the vacuum permeability. In the presence of the Rashba term in the Hamiltonian the velocity of each electron is modified: 
\begin{equation}
\bv{v}(k)=\frac{\d r}{\d t}=\frac{\i}{\hbar}\left[\mathcal{H},\bv{r}\right ]=\frac{\hbar\bv{k}}{m}+\frac{a_{\mathrm{R}}}{\hbar}(\bv{n}\times\bv{s}),
\label{e3}
\end{equation}
The second term on the r.h.s. of Eq.~(\ref{e3}) depends on spin. A precessing electron thus undergoes an oscillatory motion, which has recently been discussed as an example for {\it Zitterbewegung}.\cite{19,20,21,22}

The electron current $\bv{j}$ is obtained by averaging the spin-dependent velocity over the $\bv{k}$-space, yielding:
\begin{equation}
\bv{j}=\bv{j}_{\mathrm{d}}+\frac{ea_{\mathrm{R}}}{\hbar g\mu_{\mathrm{B}}}(\bv{n}\times\bv{M}).
\label{e4}
\end{equation}
The microwave drift (or displacement-) current, $\bv{j}_{\mathrm{d}}(\omega)=\sigma\bv{E}_1(\omega)$, described by the first term, in the low field limit is proportional to the microwave electric field $\bv{E}_1(t)=\myRe\left[\bv{E}_1(\omega)e^{-\i\omega t}\right]$   and the complex dynamic conductivity, $\sigma(\omega,\bv{H})$. The second term contains the magnetization, $\bv{M}=n_sg\mu_{\mathrm{B}}\left\langle\bv{s}\right\rangle$. Here we omit its static component $M_0$ because we investigate the microwave power absorption resulting only from spin precession. The latter causes oscillating components, and thus the second term also contributes to an oscillating current. Within the linear response approximation the amplitude of the precession is given by $M_\pm(\omega)=\chi_\pm(\omega)H_\pm(\omega)$, with the dynamic magnetic susceptibility $\chi_\pm(\omega)=\pi\gamma M_0f_\pm(\omega)$. Here the indices $\pm$  stand for the amplitudes in a rotating coordinate system and thus: $M_\pm(\omega)=\left(M_\alpha(\omega)\mp M_\beta(\omega) \right)/\sqrt{2}$, where $\alpha, \beta$  designate axes perpendicular to the direction of $\bv{H}$.
The amplitude of the oscillating field $H_\pm(\omega)$, which drives the precession is composed of the microwave magnetic field, $H_{1\pm}(\omega)$, and the S-O field, $H_{{\mathrm{R}}\pm}$. The mean value of the Rashba field is obtained by averaging Eq.~(\ref{e2}) for all electron spins:\cite{12} 
\begin{equation}
\mu_0\left\langle\bv{H}_{\mathrm{R}}\right\rangle=\frac{a_{\mathrm{R}}m^*}{eg\mu_{\mathrm{B}}\hbar n_{\mathrm{s}}}(\bv{j}_{\mathrm{d}}\times\bv{n})- \frac{a_{\mathrm{R}}m^*}{n_{\mathrm{s}}\hbar^2g^2\mu_{\mathrm{B}}^2}\left((\bv{n}\times\bv{M}_{\mathrm{s}})\times\bv{n}\right).
\label{e5}
\end{equation}
The first term results from the displacement of the Fermi sphere caused by a drift current for $\omega\tau_p<<1$ (valid also for a displacement current for $\omega\tau_p>>1$). The second one is a consequence of a change in the spin subband population which causes a non-zero mean $\bv{k}$-vector in each spin subband even without a drift current. Consequently this term contributes to the Rashba field and it is responsible \eg for the spin Hall effect.\cite{3} However, a fast precession ($\omega T_2>>1$) does not change the occupation of spin states. Therefore, the precessing magnetization does not result in any additional Rashba field and only the static (or some low frequency component) of magnetization, $\bv{M}_{\mathrm{s}}$, may occur in Eq.~(\ref{e5}).  Summarizing, the oscillating amplitude $H_{{\mathrm{R}}\pm}(\omega)$ is determined by the oscillating current amplitude, $j_{{\mathrm{d}}\pm}(\omega)$, and described by the first term of Eq.~(\ref{e5}).

\begin{figure}
\includegraphics[width=9.5cm]{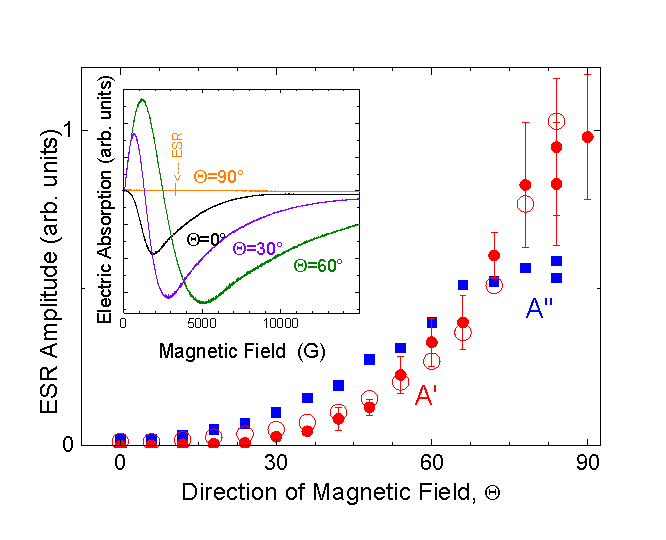}
\caption{\label{fig2}  Angular dependence of the amplitude of dispersive (dots: experimental, circles: evaluated from CR spectra) and absorption (squares, experimental) components for a sample of a moderate mobility ($\omega\tau_p=0.3$), placed in the node of $H_1$. Inset: first derivative of the CR absorption spectra for different angles $\theta$.} 
\end{figure}

Finally, the spin related, oscillating component of the total current (second term in Eq.~(\ref{e4}))  scales with $a^2_{\mathrm{R}}$, the dynamic susceptibility,  $\chi_\pm(\omega)$, and  $j_{{\mathrm{d}}\pm}(\omega)$. Therefore, both contributions to the current are proportional to $j_{{\mathrm{d}}\pm}(\omega)$ and thus to  $E_\pm(\omega)$.  But they differ in phase. The drift current $\bv{j}_{\mathrm{d}}(t)$  is shifted in phase with respect to the electric field $\bv{E}_1(t)$  by $\varphi_{jE}=\arctan{(\omega\tau_p)}$, where $\tau_p$   is the momentum relaxation time. The magnetization induced current is additionally shifted by a difference in phase between $\bv{H}_{\mathrm{R}}(t)$  and $\bv{M}(t)$. According to the solution of the Bloch equations this shift is $\varphi_{HM}=\arctan{\left((\omega-\omega_{\mathrm{L}})T_2\right)}$.

The electric power, \ie, the Joule heat, absorbed per unit area $F$ of the electron gas is given by $\d P/\d F=\left\langle\bv{j}(t)\cdot\bv{E}_1(t)\right\rangle_t$, where the current density is obtained from Eq.~(\ref{e4}). Because of the interplay of these phase shifts the spin dependent contribution is a linear combination of imaginary and real part of the magnetic susceptibility, and the amplitude of both components are described by various components of the electric conductivity tensor.	

Putting the $x$-axis into the sample plane, the in-plane component of the microwave electric field is  $E_x(\omega)=E_1(\omega)\sin{\theta}$. Then the spin dependent contribution to the Joule heat is:
\begin{widetext}
\begin{equation}
\frac{\d P_J(\omega)}{\d F}=    
\frac{C}{4}
\left\{[\mathrm{\myRe}\sigma_{xx}-\cos{\theta}\mathrm{\myIm}\sigma_{xy}]\chio_+(\omega)- 
[\mathrm{\myIm}\sigma_{xx}+\cos{\theta}\mathrm{\myRe}\sigma_{xy}]\chid_+(\omega)\right\}E_x^2(\omega),
\label{e6}
\end{equation}
\end{widetext}
with $C=a_{\mathrm{R}}^2m^*/\mu_0 g^2\mu_{\mathrm{B}}^2\hbar^2n_s$. Here we neglect contributions related to the ESR inactive component,  $\chid_-(\omega)$.
Eq.~(\ref{e6}) is the main result of this paper. It shows a spin dependent contribution to the Joule heat with a resonance at the Larmor frequency. This type of signal has both an absorptive and a dispersive component. The expressions in the square brackets are proportional to the two amplitudes, $\pri{A}_{\mathrm{J}}$  and  $\dpri{A}_{\mathrm{J}}$, respectively, of the ESR signals originating from the spin dependent motion of the electrons.  Both amplitudes can be positive or negative, depending on the electron conductivity. A negative sign shows only that the total Joule heat at resonance is {\bf partially reduced} by the spin dependent component. 

Because of the complex behavior of the electric conductivity of a high mobility 2D electron gas\cite{16} modeling is hardly possible, but for low mobility samples ($\omega\tau_p<<1$), the first term, $\pri{A}_{\mathrm{J}}$, can be approximated by the first term in the square bracket in Eq.~(\ref{e6}). In that case, the amplitude $\pri{A}_{\mathrm{J}}$  is expected to be proportional to the CR absorption at ESR, \ie, to $\d P(\omega_0)/\d F=\myRe(\sigma_{xx}(\omega_0))\left|E_x\right|^2/2$. Therefore, the angular dependence of $\pri{A}(\theta)$  can be directly compared with the experimentally measured dynamic conductivity at ESR, $\sigma_{xx}(\theta, H_{\mathrm{ESR}})$, which is obtained when the sample is placed in a node of $H_1$. The open circles in Fig.~\ref{fig2} represent the product of  $\sin^2{\theta}\cdot\myRe\sigma_{xx}(\theta,H_{\mathrm{ESR}})$. This dependence fits the experimentally observed angular dependence of $\pri{A}(\theta)$  very well, proving the proposed origin of the dispersive component in the absorption spectra.

A similar analysis of $\dpri{A}(\theta)$  is much more difficult. There are two contributions due to the energy exchange between the microwave and (i) the kinetic energy of the electrons and (ii) their magnetic energy: the former, $\dpri{A}_{\mathrm{J}}$, is described by the second term in Eq.~(\ref{e6}) and the latter by the amplitude of CI ESR,  $\dpri{A}_{\mathrm{M}}$, which is proportional to:\cite{10}
\begin{widetext}
\begin{equation}
\frac{\d P_M(\omega)}{\d F}=\frac{C}{4}\frac{m^*}{e^2n_s}\left[|\sigma_{xx}|^2+2\cos{\theta}\myIm(\sigma_{xy}^*\sigma_{xx})+ \cos^2{\theta}|\sigma_{xy}|^2 \right]\omega\dpri{\chi}_+(\omega)\cdot E^2_x(\omega).
\label{e7}
\end{equation}
\end{widetext}

The transfer to the magnetic energy (ii) has only an absorption component, proportional to  $\dpri{\chi}$.  The experimentally accessible amplitude of the absorption signal is obtained by the sum:  $\dpri{A}=\dpri{A}_{\mathrm{M}}+\dpri{A}_{\mathrm{J}}$.  Neither of them can be derived from the measured electric absorption without detailed modeling of the absorption,\cite{16} which is beyond the scope of this paper. 

\begin{figure}
\includegraphics[width=9.5cm]{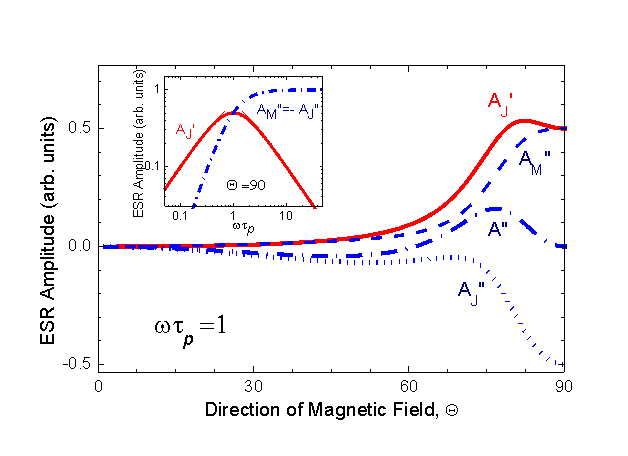}
\caption{\label{fig3}  Angular dependence of the different ESR contributions according to Eqs. (\ref{e6}--\ref{e7}) assuming the Drude model for the conductivity. The inset shows their dependence for in-plane orientation of $\bv{H}$ on  $\omega\tau_p$. Here the static magnetization is assumed to be proportional to frequency. }
\end{figure}

Therefore we discuss only a qualitative picture by approximating the electric conductivity within the Drude model. For that case we show the expected angular dependencies of the signal amplitudes in Fig.~\ref{fig3} for $\omega\tau_p=1$. The peculiarities seen close to $\theta=80\deg$  correspond to the coincidence of the spin- and cyclotron resonances for an effective electron mass of $m^*=0.2m_0$. They become more pronounced for bigger values of $\omega\tau_p$.

Within the Drude model the amplitude of the dispersive signal is expected to have a positive sign only (see Fig.~\ref{fig3}). The absorption components $\dpri{A}_{\mathrm{J}}$  and  $\dpri{A}_{\mathrm{M}}$  are of opposite sign and their sum is expected to change sign at some direction of the applied field. Altogether a quite complicated angular dependence results except for $\theta\rightarrow 90\deg$, where   $\dpri{A}_{\mathrm{J}}(90\deg)=-\dpri{A}_{\mathrm{M}}(90\deg)$, \ie, the absorption signals are expected to compensate each other.\cite{16,17}

All amplitudes resulting from the Drude model are expected to scale with  $\omega\tau_p$. The inset in Fig.~\ref{fig3} shows the dependence of the ESR signal amplitudes on $\omega\tau_p$ for $\theta=90\deg$.  Within the limit of drift conductivity (for $\omega\tau_p<<1$) the absorption amplitudes are proportional to the square of electron mobility and to $(\omega\tau_p)^2$  whereas $\pri{A}_{\mathrm{J}}$ , is proportional to  $\omega\tau_p$. Therefore $\pri{A}_{\mathrm{J}}$  dominates the absorption components. In the opposite limit of a displacement current, the absorption amplitudes become independent of frequency and momentum dissipation.  Therefore, in the infrared range, the absorption signal is expected to exceed $\pri{A}_{\mathrm{J}}$  which decreases with $(\omega\tau_p)^{-2}$. In fact, the spectra observed in the high frequency range usually exhibit an absorption type of line shape.\cite{23,24}

\section{CONCLUSION}
   
In conclusion, we presented a new effect resulting from the Rasba type of S-O coupling which manifests itself by a new type of spin resonance. The \rf Rashba field, caused by an \rf current, induces spin precession, leading to an increase of the magnetic energy. There is, however, also a component of the electric current which depends on spin orientation. Therefore the induced spin precession leads to an additional, oscillating component of the current, modifying the Joule heating. This spin dependent current is shifted in phase as compared to the classical drift current, and thus the Joule heating can increase or decrease at spin resonance, depending on the interplay of phases.  The spectral dependence of the magnetic channel is given by the absorption shape function and predicts a positive absorption only, but the line shape of the electrical channel is given by a combination of absorptive and dispersive components where the amplitude of the absorption usually has a negative sign. Consequently, the absorption signals of the two channels partially compensate each other and the resulting amplitude can be of either sign. Anyway, the absorption spectrum with the dispersive component is an attribute of the electric channel, evidencing Zitterbewegung at the Larmor frequency. Our model predicts thus a unique dependence of the resonance spectrum and of the amplitude on the experimental geometry. The experiments confirm this well, in spite of the fact, that details of the dynamic electric conductivity of small samples of high mobility 2D electron gas cannot be easily modeled.

{\bf Acknowledgments:} We would like to thank R.~Koch for a critical reading of the manuscript and F.~Sch\"affler for generously providing samples. Work supported in Austria within the FWF project 20550 of the Austrian Science Fund (Vienna) and in Poland by the MNiSW grant N~N202 1058~33.

{\footnotesize


\begin{thebibliography}{9}
\bibitem{1} E.~I.~Rashba and A.~L.~Efros, Phys. Rev. Lett. 91, 126405 (2003)
\bibitem{2} E.~I.~Rashba, Nature Physics 2, 149--150 (28 February 2006)
\bibitem{3} A recent review of these effects can be found in various contributions in ``Spin Physics in Semiconductors'', ed by: M.~Dyakonov, Springer Series in SOLID-STATE SCIENCES 157 (2008).ISSN 0171--1873
\bibitem{4} R.~Winkler: ``Spin-orbit Coupling Effects in Two-dimensional Electron and Hole Systems'' Springer Tracts in Modern Physics' (Springer, Berlin 2003). 
\bibitem{5} Z.~Wilamowski, H.~Malissa, F.~Sch\"affler, and W.~Jantsch, Phys. Rev. Lett. 98, 187203 (2007)
\bibitem{6} S~.M.~Frolov, S.~L\"uscher, W.~Yu, Y.~Ren, W.~Wegscheider, J.~A.~Folk, Nature 458, 868--871 (2009)
\bibitem{7}  K.~C.~Nowack, F.~H.~L.~Koppens, Y.~V.~Nazarov, and L.~M.~K.~Vandersypen, Science 318, 1430--1433(2007)
\bibitem{8}  D.~D.~Awschalom, M.~E.~Flatt\'e, Nature Physics 3, 153--159 (28 February 2007)
\bibitem{9}  M.~Duckheim and D.~Loss, Nature Phys. 2, 195--199 (2006)
\bibitem{10}  Z.~Wilamowski, W.~Ungier and W.~Jantsch, Phys. Rev. B78, 174423 (2008)
\bibitem{11}  For a review see: E.~I.~Rashba and V.~I.~Sheka, in Landau Level Spectroscopy, edited by G. Landwehr and E.I. Rashba, North-Holland, Amster-dam, 1991, p. 131
\bibitem{12}  Z.~Wilamowski and W.~Jantsch, Phys. Rev. B 69, 035328 (2004); and, for a review see same authors in Ref. 3 , and Physica E 10, 17--21 (2001)
\bibitem{13}  E.~Michaluk, J.~B.~Bloniarz, M.~Pabich, Z.~Wilamowski and A.~Myielski, Acta Physica Polonica 110, 263 (2006)
\bibitem{14}  M.~Schulte, J.~G.~S.~Lok, G.~Denninger, and W.~Dietsche, Phys. Rev. Lett. 94, 137601 (2005)
\bibitem{15}  S.~J.~Allen, H.~L.~St\"ormer and J.~C.~M.~Hwang, Phys. Rev. B 28, 4875 (1983)
\bibitem{16}  I.~V.~Kukushkin, J.~H.~Smet, S.~A.~Mikhailov, D.~V.~ Kulakovskii, K.~von Klitzing and W.~Wegscheider, Phys. Rev. Lett. 90, 156801 (2003)
\bibitem{17}  A.~Hochreiner, H.~Malissa, D.~Pachinger, F.~Sch\"affler, Z.~ Wilamowski and W.~Jantsch, Proc. ICPS 29, Rio de Janeiro, in press.
\bibitem{18}  W.~Ungier, W.~Jantsch, and Z.~Wilamowski, ACTA PHYSICA POL. A 112 (2): 345 (2007)
\bibitem{19}  W.~Zawadzki, Phys. Rev. B 72 085217 (2005)
\bibitem{20}   J.~ Schliemann, Daniel Loss, and R. M. Westervelt ,Phys. Rev. B 73 085323 (2006)
\bibitem{21}  U.~Z\"ulicke, R.~Winkler and J.~Bolte, Physica E 40, 1434 (2008)
\bibitem{22}  V.~Ya.~Demikhovskii, G.~M.~Maksimova, and E.~V.~Frolova, Phys. Rev. B 78, 115401 (2008)
\bibitem{23}  M.~Dobrowolska, Y.~F.~Chen, J.~K.~Furdyna, and S.~Rodriguez, Phys. Rev. Lett. 51, 134 (1983).
\bibitem{24}  The observed ESR line shape resembles the so-called Dysonian one seen in three dimensional metals. The mechanism for the latter involves the skin depth and vertical spin diffusion which do not occur in two dimensional systems.
\end{thebibliography}
\end{document}